\documentclass[aps,prl,twocolumn,showpacs,altaffillsymbol,10pt,superscriptaddress,floatfix,reprint]{revtex4-1}
\usepackage{graphicx}
\usepackage{dcolumn}
\usepackage{bm}
\usepackage{natbib}
\usepackage{amsmath}
\usepackage{amssymb}
\usepackage{CJK}
\usepackage{hyperref}
\usepackage{color}

\makeatother

\begin{document}
\newcommand{\NIST}{
National Institute of Standards and Technology,
325 Broadway, Boulder, Colorado 80305, USA}

\newcommand{\CU}{
University of Colorado, Department of Physics,
Boulder, Colorado 80309, USA}

\newcommand{\GTRI}{present address: Georgia Tech Research Institute, Atlanta, GA  30332, USA; brown171@gatech.edu}

\newcommand{\Ger}{
present address: National Physical Laboratory (NPL), Teddington, TW11 0LW, United Kingdom}

\newcommand{\Stab}{
present address: Stable Laser Systems, Boulder, Colorado, USA}

\newcommand{\uselesstown}{permanent address: Department of Physics, Korea University, 145 Anam-ro, Seongbuk-gu, Seoul 02841, South Korea}

\newcommand{\gian}{permanent address: Istituto Nazionale di Ricerca Metrologica, Strada delle Cacce 91, 10135 Torino, Italy; Politecnico di Torino, Corso duca degli Abruzzi 24, 10125 Torino, Italy}

\newcommand{\peking}{permanent address: State Key Laboratory of Advanced Optical Communication Systems and Networks, Institute of Quantum Electronics, School of Electronics Engineering and Computer Science, Peking University, Beijing 100871, China}

\newcommand{\first}{These authors contributed equally to this work.}
\title{Hyperpolarizability and operational magic wavelength in an optical lattice clock} 

\author{R. C. Brown}\thanks{\GTRI}
\affiliation{\NIST}
\thanks{Contribution of U.S. government; not subject to copyright.}
\author{N. B. Phillips}\thanks{\Stab}
\affiliation{\NIST}
\author{K. Beloy}
\affiliation{\NIST}
\author{W. F. McGrew}
\affiliation{\NIST}
\affiliation{\CU}
\author{M. Schioppo}\thanks{\Ger}
\affiliation{\NIST}
\author{R. J. Fasano}
\affiliation{\NIST}
\affiliation{\CU}
\author{G. Milani}
\thanks{\gian}
\affiliation{\NIST}
\author{X. Zhang}
\thanks{\peking}
\affiliation{\NIST}
\author{N. Hinkley}
\thanks{\Stab}
\affiliation{\NIST}
\affiliation{\CU}
\author{H. Leopardi}
\affiliation{\NIST}
\affiliation{\CU}
\author{T. H. Yoon}
\thanks{\uselesstown}
\affiliation{\NIST}
\author{D. Nicolodi}
\affiliation{\NIST}
\author{T. M. Fortier}
\affiliation{\NIST}
\author{A. D. Ludlow}\email{andrew.ludlow@nist.gov}
\affiliation{\NIST}
%
\begin{abstract}
Optical clocks benefit from tight atomic confinement enabling extended interrogation times as well as Doppler- and recoil-free operation.
However, these benefits come at the cost of frequency shifts that, if not properly controlled, may degrade clock accuracy.
Numerous theoretical studies have predicted optical lattice clock frequency shifts that scale nonlinearly with trap depth.
To experimentally observe and constrain these shifts in an $^{171}$Yb optical lattice clock, we construct a lattice enhancement cavity that exaggerates the light shifts.
We observe an atomic temperature that is proportional to the optical trap depth, fundamentally altering the scaling of trap-induced light shifts and simplifying their parametrization.
We identify an ``operational" magic wavelength where frequency shifts are insensitive to changes in trap depth.
These measurements and scaling analysis constitute an essential systematic characterization for clock operation at the $10^{-18}$ level and beyond.
\end{abstract}
\date{\today}
\maketitle

Optical dipole trapping has risen from theory~\cite{Letokhov1968} to establish itself as a workhorse experimental technique in numerous contexts~\cite{grimm2000odt,bloch2008RMP,kimble2008quantumintertnet,Light-ShiftCompensated_PRL}.
Despite the fact that dipole trapping is achieved by inducing large light shifts, it has found prominence in quantum metrology and precision measurements.
The concept of magic wavelength trapping resolves this apparent contradiction by inducing \emph{identical} shifts on two atomic states of interest~\cite{ye2008state_insensitive_trap}.
In an optical clock, the energy difference of these states gives the frequency reference that serves as the timebase.
The magic wavelength allows optical lattice clocks~\cite{OLC_Takamoto2005} to realize the unperturbed atomic transition frequency while maintaining the experimental benefits of trapped systems.
Magic wavelength trapping has found applications far beyond atomic clocks including: cavity QED~\cite{mckeever2003}, ultracold molecules~\cite{danzl2010ultracold} and Rydberg gases~\cite{RydbergMagic_PRA_2015}, atomic qubits~\cite{LundbladPRA_magicRb,magic_intensity_optical_traps2016}, laser cooling~\cite{MagicLaserCooling}, and quantum simulation~\cite{scazza2014SUN,cappellini2014SUN}.

Magic wavelength optical lattices have enabled optical clocks to achieve unprecedented levels of performance, with fractional frequency instability approaching $1\times10^{-18}$~\cite{HinkleyScience2013,bloomNature2014optical,Ushijima2015cryogenic,NicholsonNatCom2015,Schioppo2016} and total systematic uncertainty in the $10^{-18}$ range~\cite{bloomNature2014optical,Ushijima2015cryogenic,NicholsonNatCom2015}.
Consequently, optical clocks become sensitive tools to measure the gravitational red shift and geopotential~\cite{ashby2009Gravity,schiller2009Gravity,chou2010relativity,Geopotential2016NatPho}, search for dark matter \cite{derevianko2014DarkMatterClocks,Arvanitaki2015DarkMatterClocks,DarkMatterOpticalAtomicClock2016}, constrain physics beyond the Standard Model \cite{HuntemannPRLalphadot,2016arXiv1NewPhysicsIS,2016arXivNewPhysicsIS}, improve very long baseline interferometry~\cite{Normile2011VLBI}, and ultimately redefine the second~\cite{Riehle2015OpticalSecond}.  
However, at these performance levels, the concept of magic wavelength confinement breaks down \cite{OLC_proposal_2003,TaichenachevPRL2008}.
Higher-order couplings, including magnetic dipole~(M1), electric quadrupole~(E2), and hyperpolarizability, prevent a complete cancellation of the light shifts between clock states, introducing shifts that scale nonlinearly with trap depth.

In an $^{171}$Yb optical lattice clock, we measure nonlinear light shifts, offering improved determinations of the hyperpolarizability and lattice magic frequency~$\nu_{\mathrm{magic}}$~\cite{BarberPRL2008,LatticeClockLemkePRL2009,Nemitz2016,INRIM2016}.
Theoretical studies suggest that these higher-order light shifts yield lattice-band-dependent effects~\cite{TaichenachevPRL2008,InsensitivetoMotion_PRL_2009,OvsiannikovPRA2013,KatoriPRA2015} which vary with atomic temperature, complicating characterization of the light shift and its appropriate extrapolation to zero.
In this Letter, we extend the theory and experimentally study these temperature-dependent effects.
Doing so reveals a simplification in the shift's functional form, achieving $1.2 \times 10^{-18}$ clock shift uncertainty.
The nonlinear shifts offer an experimental benefit in the form of `operational magic wavelength' behavior - where the polarizability can be tuned, with laser frequency, to partially compensate the hyperpolarizability and yield linear shift insensitivity to trap depth.
These measurements and analysis are relevant for other atomic species, including $^{87}$Sr, where the role of hyperpolarizability for accurate characterization of lattice light shifts differs between studies~\cite{SyrteHyppol2006,WestergaardPRL2011,leTargatNatCom2013,falke2014strontium,Ushijima2015cryogenic,NicholsonNatCom2015}.

\begin{figure}
\centering\includegraphics [width=3.2in,angle=0] {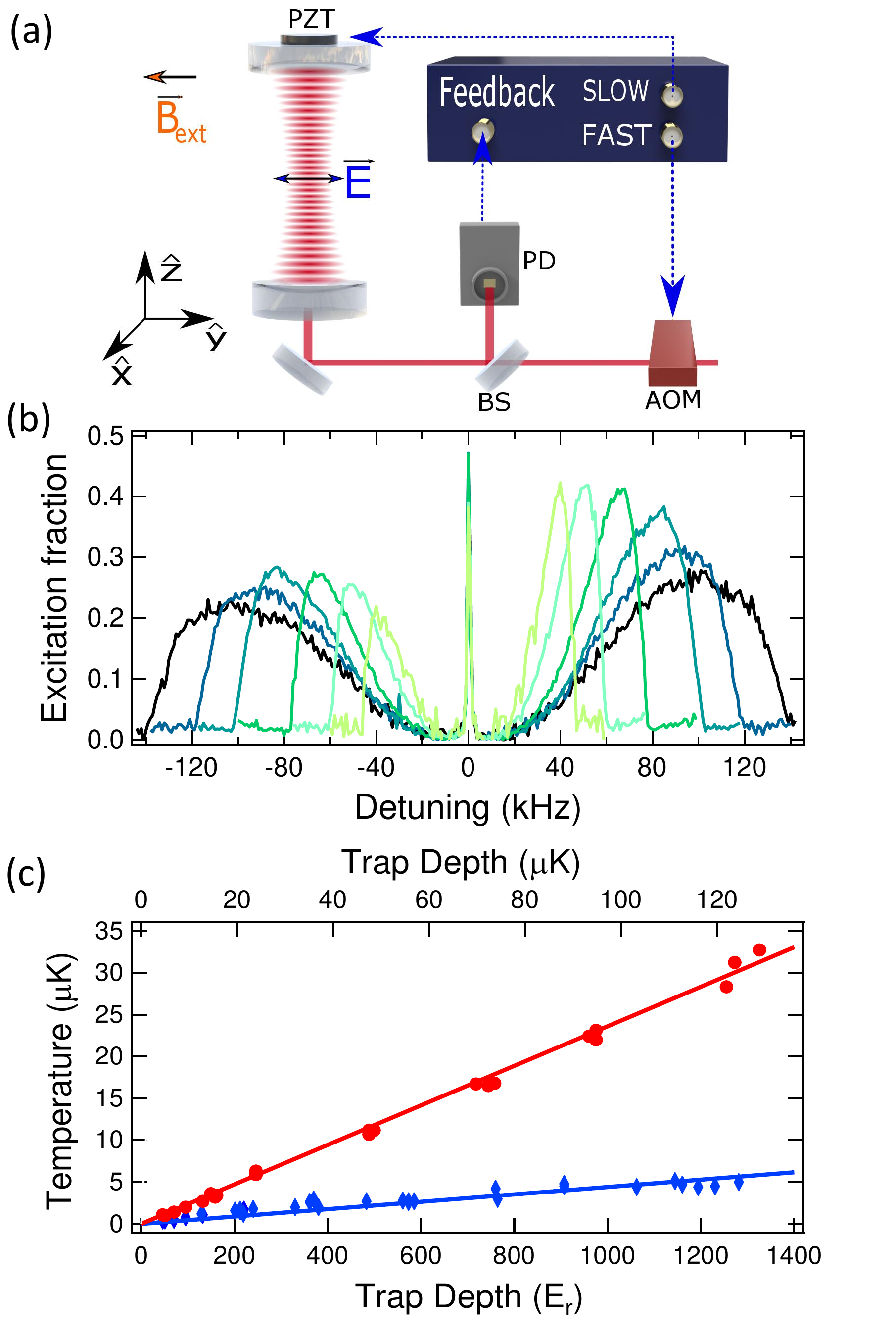}
\caption{(a) Schematic of the vertically-oriented lattice build up cavity, with out-of-vacuum mirrors. PD photodiode, BS beamsplitter, AOM acousto-optical modulator, PZT piezo-electric transducer.  (b) Sideband spectra for multiple trap depths from 150~$E_r$ (light green trace) to 1260~$E_r$ (black trace), shown as the measured excited state ($^3P_0$) fraction versus laser detuning from the $^1S_0$-$^3P_0$ clock transition frequency (c) Longitudinal temperatures, which characterize the Boltzmann distribution of atomic population among the lattice bands, are extracted from sideband spectra over a range of trap depth.  The red trace corresponds to normal operating conditions, while the blue trace incorporates an additional step of sideband cooling.}
\label{fig:TempDep}
\end{figure}

The dominant optical-trap AC Stark effect is from electric dipole polarizability ($\alpha_{E1}$), giving a shift that scales to leading order with trap depth.
The differential shift of the clock transition is eliminated at the magic frequency~\cite{OLC_proposal_2003}.
Higher multipolarizabilities from magnetic dipole and electric quadrupole contributions (denoted here as $\alpha_{M1E2}$) yield much smaller shifts.
The hyperpolarizability ($\beta$) shift accounts for electric dipole effects that are fourth order in the electric field.
In general, the frequency shift on the clock transition, $\delta \nu_{\mathrm{clock}}$, is:
\begin{equation}
\frac{\delta \nu_{\mathrm{clock}}}{\nu_{\mathrm{clock}}}=
-U\,\Delta \alpha_{E1}'\,X_\mathbf{n}
-U\,\Delta \alpha_{M1E2}'\,Y_\mathbf{n}
-U^2\,\Delta \beta'\,Z_\mathbf{n},
\label{Eq:new1}
\end{equation}
where all quantities appearing on the right-hand-side are dimensionless (see Supplemental).
Here, $\Delta$ denotes a difference in a quantity between clock states, and $\Delta \alpha_{E1}' = \Delta \alpha_{E1} E_r/ \alpha_{E1}(\nu_{\mathrm{magic}}) h\nu_{\mathrm{clock}}$, $\Delta \alpha_{M1E2}' = \Delta \alpha_{M1E2} E_r/ \alpha_{E1}(\nu_{\mathrm{magic}}) h\nu_{\mathrm{clock}}$, $\Delta \beta' = \Delta \beta {E_r}^2/\alpha_{E1}(\nu_{\mathrm{magic}})^2 h\nu_{\mathrm{clock}}$.  $X_\mathbf{n}$, $Y_\mathbf{n}$, and $Z_\mathbf{n}$ represent expectation values of the spatial portion of the trapping potential, $\mathcal{U}(z,\rho)=\exp\left(-2\rho^2/{w_0}^2\right)\cos^2\left(kz\right)$, for motional state $\mathbf{n}$ with $1/e^2$ lattice-beam-intensity radius $w_0$; $X_\mathbf{n}\equiv\left\langle\mathbf{n}\left|\mathcal{U}(z,\rho)\right|\!\mathbf{n}\right\rangle,
Y_\mathbf{n}\equiv\left\langle\mathbf{n}\left|\mathcal{U}(z+\pi/(2k),\rho)\right|\!\mathbf{n}\right\rangle, Z_\mathbf{n}\equiv\left\langle\mathbf{n}\left|
\mathcal{U}(z,\rho)^2\right|\!\mathbf{n}\right\rangle$.  
$U$, which is proportional to lattice intensity, is the dimensionless ratio of trap depth to recoil energy $E_r=\frac{\hbar^2 k^2}{2 m}$, where $k=2 \pi \nu_l/ c$ for lattice frequency~$\nu_l$, $h=2\pi \hbar$ is Planck's constant, $c$ is the speed of light, and $m$ is the mass of $^{171}$Yb.  

Here, we extend the perturbative treatment in the harmonic motional-state basis~\cite{KatoriPRA2015} to consider not only longitudinal confinement along the 1-D optical lattice, but also transverse optical confinement.
The resulting lattice-induced shift for an atom in longitudinal lattice band $n_z$ and transverse motional state $n_{\rho}=n_x+n_y$ is:
\begin{multline}
\frac{\delta \nu_{\mathrm{clock}}}{\nu_{\mathrm{clock}}} = n_{5} \Delta \alpha_{M1E2}'
+ [(n_{1} +n_{2}) \Delta \alpha_{E1}' - n_{1} \Delta \alpha_{M1E2}'] U^{\frac{1}{2}}\\
-[\Delta \alpha_{E1}' + (n_{3}+n_{4}+4n_{5}) \Delta \beta'] U
+[2 \Delta \beta' (n_{1}+n_{2})] U^{\frac{3}{2}}
-\Delta \beta' U^2.
\label{eq:fullfreqshift}
\end{multline}   
This treatment yields a $U^{1/2}$ scaling originating from $\alpha_{M1E2}$~\cite{TaichenachevPRL2008,InsensitivetoMotion_PRL_2009} and a $U^{3/2}$ scaling originating from $\beta$~\cite{OvsiannikovPRA2013} and now includes contributions from both the transverse and longitudinal motional quantum numbers: $n_{1}=(n_z+1/2)$, $n_{2}=\frac{\sqrt{2}}{kw_0} (n_{\rho}+1)$, $n_{3}=\frac{3}{2} (n_z^2+n_z+1/2)$, $n_{4}= \frac{8}{3k^2w_0^2}(n_{\rho}^2+2 n_{\rho}+3/2)$, and $n_{5}=\frac{1}{\sqrt{2}kw_0}(n_z+1/2)(n_{\rho}+1)$.

Since measurements cannot be made at zero trap depth, extrapolation to the unperturbed clock transition frequency at $U=0$ is required.
For shallow traps with cold low-density atomic samples, an extrapolation linear in $U$ has generally been considered sufficient to determine the magic wavelength and unperturbed atomic frequency at the $10^{-17}$ level of clock uncertainty.
However, as the required accuracy increases, the higher order terms in Eq.~(\ref{eq:fullfreqshift}) cannot, in general, be neglected.
The added fit parameters from each $U$-dependent term place a heavy statistical burden on the measurement in order to reach the desired level of uncertainty.
Furthermore, the inclusion of these higher-order terms introduces contributions dependent on the thermally averaged $\langle n \rangle$.
In order to meaningfully apply Eq.~(\ref{eq:fullfreqshift}) to experimental data, the $\langle n \rangle$ must be characterized over the range of $U$ measured.

To experimentally observe light shifts in an $^{171}$Yb optical lattice clock ~\cite{HinkleyScience2013}, we use a power enhancement cavity (finesse $\approx$ 200 at $\nu_l$, transparent at $\nu_{\mathrm{clock}}$) to form the optical lattice, Fig.~\ref{fig:TempDep}(a), enabling trap depths $>$$20\times$ our anticipated operational depth.
A relatively large lattice beam radius ($170$ $\mu$m) in the transverse plane enables high trapped atom number with relatively low atomic density and thus small density-dependent collisional shifts.
The cavity orientation along gravity suppresses resonant tunneling between lattice sites~\cite{vert,LemondeWolfPRA2005}.
To ensure that the optical lattice has no significant residual circular polarization (e.g.~vacuum window birefringence), the difference frequency between $\pi$-transitions from both $m_F = \pm 1/2$~\cite{tensor} states is measured for all $U$.
Residual circular polarization would cause a $U$-dependent vector AC Stark shift in the observed splitting.
No such dependence is observed, allowing us to constrain lattice ellipticity below 0.6$\%$.
Using the vector AC Stark splitting as an in-situ atomic sensor of optical lattice polarization allows us to exclude polarization-dependent variations in the observed hyperpolarizability effect \cite{LatticeClockLemkePRL2009}.
The lattice laser frequency is stabilized, over the course of a measurement, to a cavity with a typical drift of $\lesssim\!100$ kHz per day.
The absolute lattice laser frequency was measured to within $\approx$10~kHz using a referenced Ti:sapphire optical frequency comb~\cite{fortier2011,fortier2006octave}.

Atomic temperature in both the longitudinal and transverse lattice dimensions, as well as the magnitude of $U$, is assessed for all clock shift measurements via motional sideband spectroscopy, Fig.~\ref{fig:TempDep}(b)~\cite{BabyBlattPRA2009}.
We observe that the temperature scales predominantly linear in $U$, Fig.~\ref{fig:TempDep}(c).
We attribute this linear scaling to the interplay of lattice induced light shifts on the ${^1\mathrm{S}_0} \rightarrow {^3\mathrm{P}_1}$ cooling transition and the atomic kinetic energy cutoff imposed by the finite lattice depth.
The linear scaling of temperature with $U$ has important consequences: for our observed ratio of temperature to trap depth, the following lowest-order approximations hold:  $\langle n_{1}\rangle,\langle n_{2}\rangle\!\propto\!\sqrt{U}$ and $\langle n_{3}\rangle,\langle n_{4}\rangle,\langle n_{5}\rangle\!\propto\!U$.
Under these conditions, Eq.~(\ref{eq:fullfreqshift})~ can be reduced to:
\begin{equation}
\delta\nu_{\mathrm{clock}}/\nu_{\mathrm{clock}}=-\alpha^{*}U-\beta^{*}U^{2},
\label{Eq:dvavg}
\end{equation}
with $U$-independent finite-temperature polarizabilities $\alpha^{*}$ and $\beta^{*}$ (see Supplemental).

\begin{figure}
\centering\includegraphics [width=3.6in,angle=0] {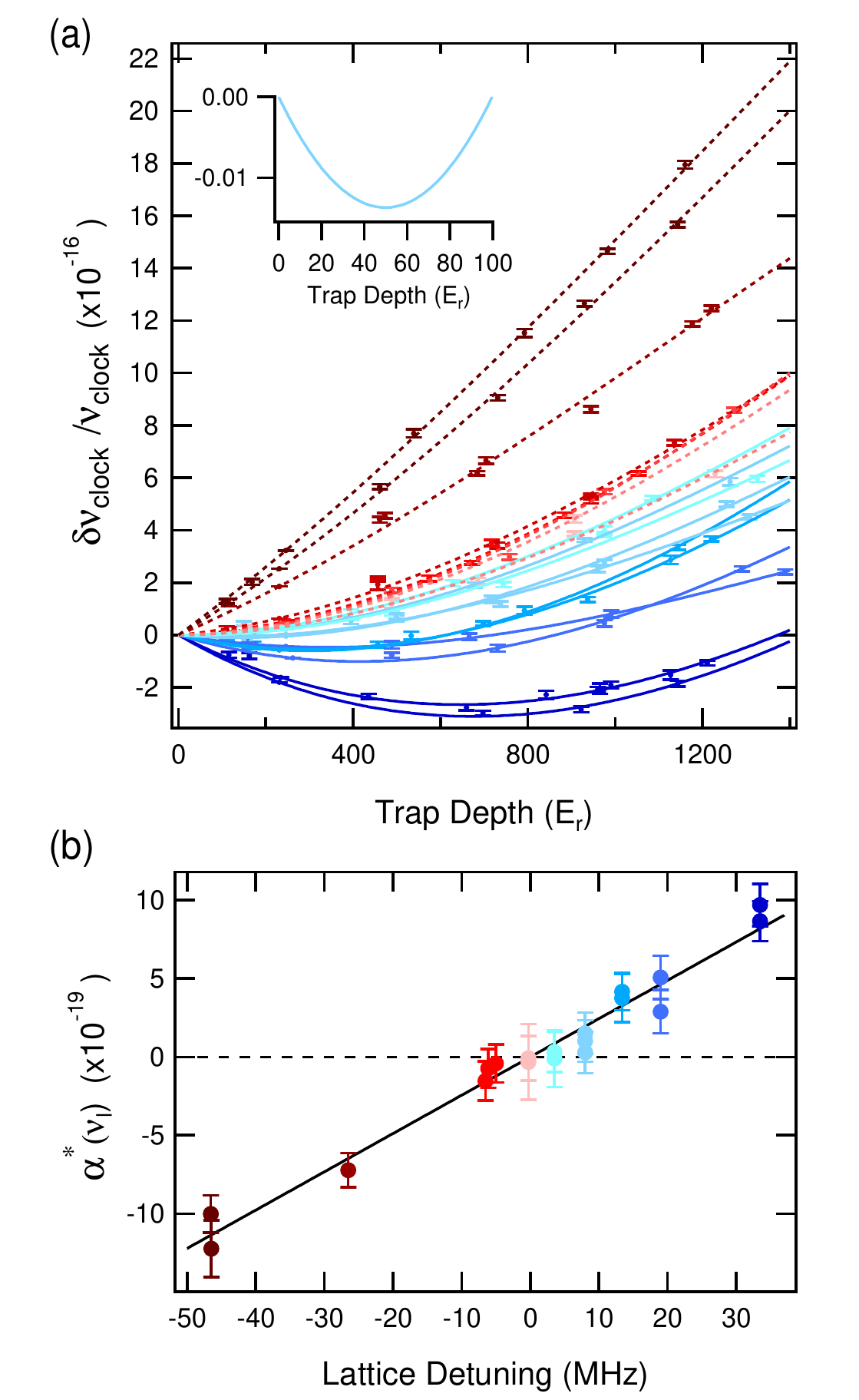}
\caption{(a) Clock shifts as a function of lattice depth.  Colored traces represent data sets with distinct detunings of $\nu_l$ from $\nu_{\mathrm{zero}}$ from $\approx-50$~MHz (dark red) to $\approx30$~MHz (dark blue). This color scheme is quantified in Fig.~2(b).  Inset) At the operational magic wavelength for a 50~$E_r$ lattice depth, a 10$\%$ change in trap depth creates a $1 \times 10^{-19}$ change in $\delta\nu_{\mathrm{clock}}/\nu_{\mathrm{clock}}$.
(b) Linear coefficients from the global fit, primarily proportional to $\Delta \alpha_{E1}$, as a function of lattice laser detuning from $\nu_{\mathrm{zero}}$ . This data is corrected for measured density shifts but not for calculated M1/E2 effects.}
\label{fig:ClockShft}
\end{figure}
Intensity dependent light shifts were measured with interleaved comparisons of the frequency shift between test- and reference-lattice depth clock configurations, as in Ref.~\cite{LatticeClockLemkePRL2009}.
Sideband spectra were taken directly before or after interleaved clock comparison.
The density shift was independently measured as a function of trap depth to apply small ($<4 \times 10^{-18}$) corrections to the measured light-shift data, with minimal impact on the deduced magic wavelength.
For a given lattice frequency, clock shifts were measured as a function of trap depth, Fig.~\ref{fig:ClockShft}(a).
Each color represents data sets with a distinct $\nu_l$.
The uncertainties in $\delta\nu_{\mathrm{clock}}/\nu_{\mathrm{clock}}$ are the total Allan deviation at the end of each data run~($\approx 1 \times 10^{-17}$).

We analyze the experimental data in Fig.~\ref{fig:ClockShft}(a) by fitting each data set to a modified form of Eq.~(\ref{Eq:dvavg})~(plus a constant term to account for the $U\neq0$ reference condition).  In principle, a fit with a single quadratic coefficient could be justified because hyperpolarizability has negligible lattice frequency dependence in the vicinity of the magic wavelength.  Nevertheless, it is possible for $\Delta \alpha_{E1}$ effects to couple to $\beta^{*}$, giving it dependence on lattice frequency.  This situation can arise, for example, from atomic temperature that scales nonlinear in the trap depth. Therefore, we perform fits with and without a global $\beta^{*}$, with both methods yielding a mean value of $\beta^{*} = -5.5(2)\times 10^{-22}$~\cite{error}.  $\beta^{*} < \Delta \beta'$ due to the finite temperature of the system; atoms in higher motional states are more spatially delocalized and thus experience lower average lattice laser intensity.  Nonlinear scaling of the atomic temperature can have other important consequences, such as light shifts with additional $U$-dependencies that must be included in Eq.~(\ref{Eq:dvavg}) for high accuracy shift determination.  Because we have observed a residual quadratic dependence of the transverse atomic temperature versus trap depth, we also allow for a $U^3$-dependent fit term (see Supplemental).  The linear coefficients, $\alpha^{*}$, extracted from the fits to data in Fig.~\ref{fig:ClockShft}(a), are shown in Fig.~\ref{fig:ClockShft}(b).
These coefficients scale linearly with the lattice detuning and are parameterized as $\alpha^{*}(\nu_l) =(\partial\alpha^{*}/\partial\nu_l)\times (\nu_l-\nu_{\mathrm{zero}})$.
Fitting to this functional form, we find ${\partial\alpha^{*}/\partial\nu_l = 2.46(10)\times 10^{-20} \frac{1}{\mathrm{MHz}}}$ and that the linear shift vanishes at $\nu_{\mathrm{zero}}= 394,798,267(1)$~MHz.
Using a second independent atomic system with similar experimental conditions, we observe consistent values of $\partial\alpha^{*}/\partial\nu_l$, $\beta^{*}$, and $\nu_{\mathrm{zero}}$ between the two systems.
For anticipated clock operation with a trap depth of $50$ $E_r$, our determinations of $\alpha^{*}$ and $\beta^{*}$ are sufficient for $10^{-18}$ uncertainty.

By inspection of Eq.~(\ref{eq:fullfreqshift}), and with $\langle n_{1}\rangle,\langle n_{2}\rangle \propto \sqrt{U}$ and $\langle n_{5}\rangle \propto U$, we see that both E1 and M1/E2 frequency shifts scale linearly with $U$.
The dominant effect of M1/E2 contributions is to thus move the observed zero value of the linear shift away from the lattice frequency where $\Delta \alpha_{E1} =0$, $\nu_{\mathrm{zero}}=\nu_{\mathrm{magic}}-\nu_{M1E2}$.
To estimate the effect, we perform a configuration interaction plus many-body perturbation theory calculation~\cite{DzubaJPhysB2010} and determine $ \Delta \alpha_{M1E2}'=4(4) \times 10^{-8} \left( \frac{E_r}{h\nu_{\mathrm{clock}}} \right)$ corresponding to $\nu_{M1E2}\approx -400$~kHz.  
This result follows from the partial cancellation of larger terms, yielding a large relative uncertainty.
Although $\nu_{\mathrm{magic}}$ can be deduced from our experimentally measured $\nu_{\mathrm{zero}}$ and theoretically calculated $\nu_{M1E2}$, we emphasize that $\nu_{\mathrm{zero}}$ represents an experimentally relevant quantity to zero all linear shifts in Eq.~(\ref{Eq:dvavg}).

\begin{figure}
\centering\includegraphics [width=3.2in,angle=0] {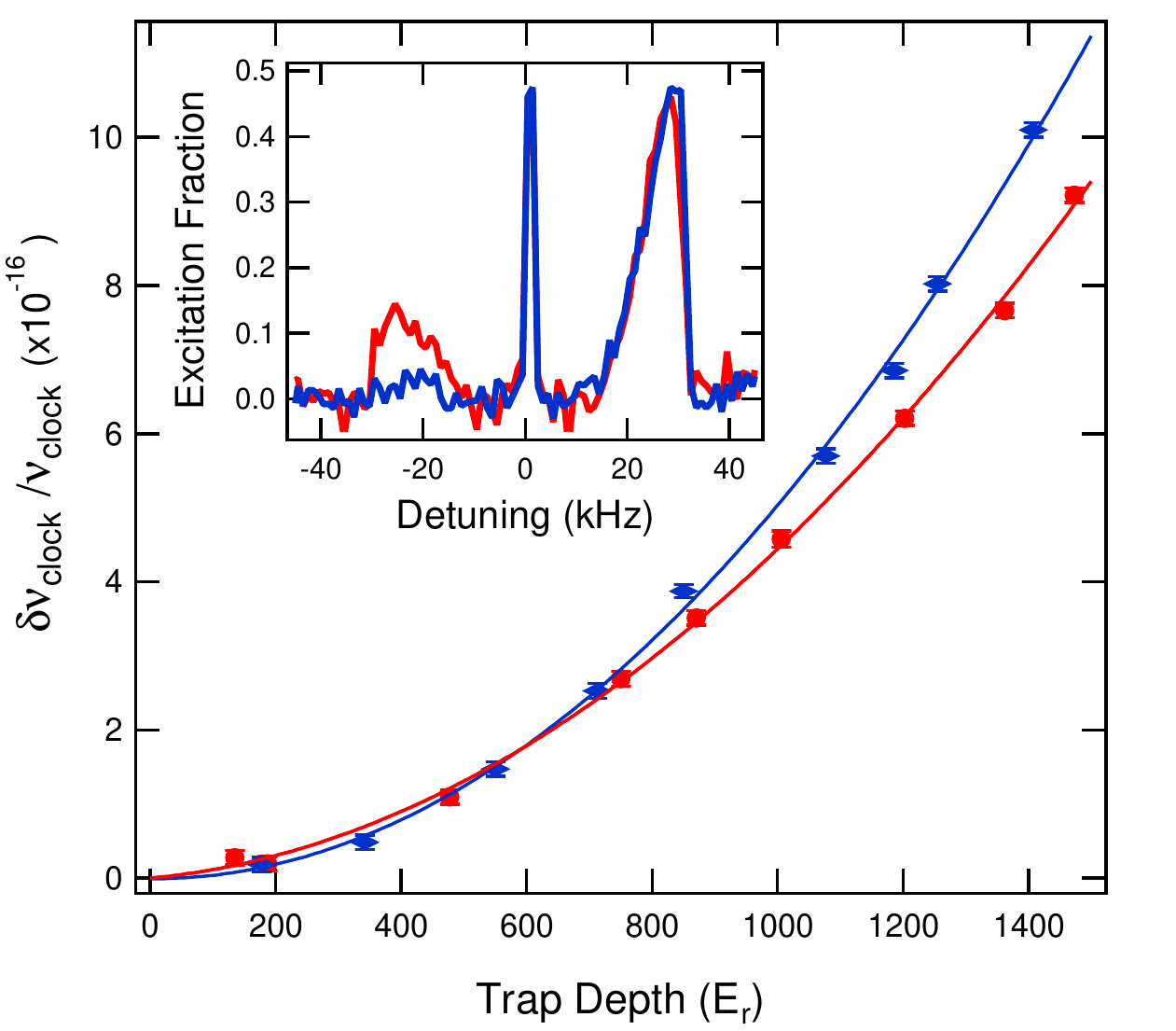}
\caption{(Color online) To experimentally explore the role of finite temperature effects, we measure $\delta\nu_{\mathrm{clock}}/\nu_{\mathrm{clock}}$ near $\nu_{\mathrm{zero}}$ both with (dark gray(blue) data) and without (light gray(red) data) sideband cooling.  Since the cooler atoms are more localized in the high-intensity portion of the lattice, they experience a larger shift originating from the hyperpolarizability. The inset shows representative sideband traces.}
\label{fig:ClockShiftSBcool}
\end{figure}

To highlight the important role of atomic temperature, we measure lattice light shifts under two distinct thermal conditions.
Figure~\ref{fig:ClockShiftSBcool} displays the light shift versus trap depth with and without an additional stage of cooling along the lattice axis dimension, using quenched sideband cooling on the ultra-narrow ${^1\mathrm{S}_0} \rightarrow {^3\mathrm{P}_0}$ clock transition \cite{CurtisQuenched2001,Nemitz2016}.  As seen in Fig.~\ref{fig:TempDep}(c), the sideband cooling reduces the longitudinal temperature by a factor of $\geq 6$, ranging from just 400~nK to $5~\mu$K and with a predominantly linear dependence on $U$.
In Fig.~\ref{fig:ClockShiftSBcool}, the observed shifts are larger in the cooled case, since the near-unity population in the ground lattice band experiences the highest lattice laser intensity. The measured hyperpolarizability effect in the sideband-cooled case increases $\beta^{*}$ by $12(5)\%$.
This change in $\beta^{*}$ introduced by cooling just one dimension underscores the importance of characterizing thermal effects on lattice shifts.

Using the preceding expressions and taking into account thermal effects, we translate the measured $\beta^*$ and $\alpha^*$ to the respective atomic properties $\Delta\beta^\prime\approx -10 \times 10^{-22}$ and $\partial\Delta\alpha_{E1}^\prime/\partial \nu_l \approx 4 \times 10^{-20} \frac{1}{\mathrm{MHz}}$.
Alternatively, known lifetime and polarizability data can be used to calculate $\partial\Delta\alpha_{E1}^\prime/\partial\nu_l= 4.5(3) \times 10^{-20} \frac{1}{\mathrm{MHz}}$.
While agreement between theory and experiment is reassuring, the perturbative treatment does not fully account for anharmonic and cross-dimensional effects relevant for higher-lying motional states.
We have developed more sophisticated models to evaluate Eq.~(\ref{Eq:new1}) accounting for these effects~\cite{Beloy2019}.
Importantly, we find a key behavior is maintained in more refined analyses: given a linear relationship between temperature and depth, the clock shift is well-approximated by Eq.~(\ref{Eq:dvavg}) with $\alpha^*$ and $\beta^*$ being independent of depth.

The fitted parameters enable us to identify a $U$-dependent \emph{operational} magic frequency.  Neglecting any residual $U^3$ shift dependence or $\beta^{*}$ detuning dependence, ${\nu_{\mathrm{opmagic}} \equiv (-2\beta^{*}U)/(\partial\alpha^{*}/\partial\nu_l)+\nu_{\mathrm{zero}}}$.
At this value of $\nu_l$ and corresponding $U$, a negative linear light shift partially cancels the positive hyperpolarizability shift, yielding a shift with first-order insensitivity to fluctuations in $U$.
Solving for a trap depth at 50~$E_r$, the measurements in Fig.~\ref{fig:ClockShft} indicate an operational magic wavelength of 2.2(1)~MHz above~$\nu_{\mathrm{zero}}$.
Although typically controlled at the 1$\%$ level, a 10$\%$ change in trap depth creates a $< 1 \times 10^{-19}$ change in $\delta\nu_{\mathrm{clock}}/\nu_{\mathrm{clock}}$. 
This parameter regime is shown as an inset in Fig.~\ref{fig:ClockShft}(a).

While the combination of hyperpolarizability and lattice detuning are useful for achieving operational magic wavelengths, they can also obscure determination of $\nu_{\mathrm{zero}}$ and $\nu_{\mathrm{clock}}$ when deduced from measurements experimentally limited to a restricted range of $U$.
In the simplest case, one can mistake a local minimum for a flat line leading to extrapolation errors in $\nu_{\mathrm{clock}}$ and incorrect determinations of $\nu_{\mathrm{zero}}$.
Consider our measured parameters ($\beta^{*} = -5.5(2)\times10^{-22}$, $\partial\alpha^{*}/\partial\nu_l = 2.46(10)\times10^{-20} \frac{1}{\mathrm{MHz}}$) and experimental shift uncertainties~$\pm1\times10^{-17}$.
For a measurement range limited from 100 to 300~$E_r$, variation of lattice light shifts would be~$< 6\times10^{-18}$ at a detuning of 8.9 MHz from $\nu_{\mathrm{zero}}$ (the operational magic wavelength for the middle of the measurement interval: 200$E_r$).
At this detuning, the clock shift would appear independent of $U$, giving the illusion of magic wavelength operation and making it statistically challenging to resolve hyperpolarizability or non-magic linear shifts~\cite{Ftest}.
Linearly extrapolating to $U=0$, errors in $\delta \nu_{\mathrm{clock}}/\nu_{\mathrm{clock}}$ of $2 \times 10^{-17}$ and a corresponding error in $\nu_{\mathrm{zero}}$ of 8.9 MHz could result.
Such a difficulty in resolving hyperpolarizability and the resulting error in the light shift determination is general for all lattice laser frequencies (not restricted to $\nu_{\mathrm{opmagic}}$) and may apply to other atomic species.
The case of $^{87}$Sr is notable, due to previous measurements and disagreement about the role of hyperpolarizability~\cite{SyrteHyppol2006,WestergaardPRL2011,leTargatNatCom2013,falke2014strontium,Ushijima2015cryogenic,NicholsonNatCom2015}.
While the scaling of atomic temperature with trap depth has not been fully considered, experimental parameters have been reported for strontium~(${\Delta\beta'} = -10(3)\times10^{-22}$~\cite{leTargatNatCom2013}, ${\Delta\beta'} = -7(7)\times10^{-22}$~\cite{NicholsonNatCom2015}, and $\partial{\Delta \alpha'_{E1}}/\partial\nu_l = 3.6(3) \times10^{-20} \frac{1}{\mathrm{MHz}}$~\cite{SyrtePolarizabilitiesPRA2015}).
A similar analysis to that above finds linear versus nonlinear extrapolations over the same limited range of $U$ leads to differences in the shift determination $\delta \nu_{\mathrm{clock}}/\nu_{\mathrm{clock}}$ up to $(2-4) \times 10^{-17}$.
It seems that the role of nonlinear extrapolations in $^{87}$Sr will hinge on developing consensus on the magnitude of $\beta^{*}$, including proper accounting of the temperature scaling with $U$.
Furthermore, this consideration can guide ongoing work in Mg~\cite{MgClockish}, Hg~\cite{HgClock}, and Cd~\cite{Cd_effort}.

In conclusion, we have precisely characterized optical lattice induced light shifts including nonlinear hyperpolarizability effects.
Our measurements highlight the importance of finite temperature effects at $10^{-18}$ fractional frequency accuracy.
We have also experimentally demonstrated a metrologically useful regime, the operational magic wavelength, where changes in light shifts can be minimized as the trap depth changes.
Furthermore, by implementing quenched sideband cooling along the 1-D lattice axis, tunneling related shifts are suppressed, while somewhat warmer transverse temperatures reduce overall lattice light shifts.
These measurements further lay the framework for controlling lattice light shifts at the $10^{-19}$ level.

\begin{acknowledgments}
This work was supported by NIST, NASA Fundamental Physics, and DARPA QuASAR.  R.C.B. acknowledges support from the NRC RAP.
We appreciate absolute frequency comb measurements by  F. Quinlan, useful discussions with C.~Oates, D.~Hume, and G.~Hoth, and technical assistance from J.~Sherman.

R.C.B. and N.B.P. contributed equally to this work.
\end{acknowledgments}

\begin{thebibliography}{63}
\bibitem[{ver()}]{vert}
\bibinfo{note}{{The lattice is $\approx1.5^{\circ}$ off of vertical to avoid
  spurious reflections from view ports.}}
\bibitem[{ten()}]{tensor}
\bibinfo{note}{{Since $^{171}$Yb has a nuclear spin of $I=1/2$, there is no
  tensor shift.}}
\bibitem[{err()}]{error}
\bibinfo{note}{{For nonlinear fits with $\chi^2>1$, not all fitting routines by
  default return the same standard errors. In the case of $\chi^2>1$, we report
  uncertainty multiplied by the square root of the reduced $\chi^2$.}}
\bibitem[{Fte()}]{Ftest}
\bibinfo{note}{{A F-test at the 5$\%$ level for higher order terms would favor
  the null hypothesis of a lower order fit.}}
\end{thebibliography}

\section{Supplemental Material}
\subsection{Lattice potential and Stark shifts}

We start with the potential
\begin{align}
V(\rho,z)=&{}
-\left(\frac{\mathcal{E}_0}{2}\right)^2\alpha_{E1}\exp\left(-\frac{2\rho^2}{w_0^2}\right)\cos^2\left(kz\right)
\nonumber\\&
-\left(\frac{\mathcal{E}_0}{2}\right)^2\alpha_{M1E2}\exp\left(-\frac{2\rho^2}{w_0^2}\right)\sin^2\left(kz\right)
\label{Eq:Vfull}\\&
-\left(\frac{\mathcal{E}_0}{2}\right)^4\beta\exp\left(-\frac{4\rho^2}{w_0^2}\right)\cos^4\left(kz\right),
\nonumber
\end{align}
where $\mathcal{E}_0$ is the electric field amplitude along the lattice axis at the location of the electric field anti-nodes. The first and last terms here account for $E1$ effects second and fourth order in the field amplitude, respectively, while the middle term encapsulates $M1$ and $E2$ effects second order in the field amplitude. All omitted contributions (higher order in the field amplitude or higher multipolarity) are highly suppressed relative to the terms included here. The $E1$ polarizability $\alpha_{E1}$, the $M1/E2$ polarizability $\alpha_{M1E2}$, and the hyperpolarizability $\beta$ depend on the clock state. The potential is dominated by the first term, proportional to $\alpha_{E1}$. Critically, the lattice laser frequency is chosen in the vicinity of the magic frequency, where $\alpha_{E1}$ is identical for the two clock states. We denote the value of $\alpha_{E1}$ at the magic frequency as $\alpha_{E1}\left(\nu_\mathrm{magic}\right)$. Note that regardless of the lattice laser frequency ultimately chosen for operation, the quantity $\alpha_{E1}\left(\nu_\mathrm{magic}\right)$ is unambiguously defined. From the preceding arguments, a fair approximation to the potential, appropriate for either clock state, is
\begin{equation*}
V_0(\rho,z)=
-\left(\frac{\mathcal{E}_0}{2}\right)^2\alpha_{E1}\left(\nu_\mathrm{magic}\right)\exp\left(-\frac{2\rho^2}{w_0^2}\right)\cos^2\left(kz\right).
\end{equation*}
We refer to this potential as the ``magic'' potential. The motional states associated with this potential are denoted as $\left|\mathbf{n}\right\rangle$.

The depth of the magic potential is given by $\left(\mathcal{E}_0/2\right)^2\alpha_{E1}\left(\nu_\mathrm{magic}\right)$. Formally, this not equivalent to the depth of the full potential, Eq.~(\ref{Eq:Vfull}), for either clock state. Considering the non-magic effects contribute fractionally $\lesssim\!10^{-7}$ to the depth, in many contexts the distinction is completely irrelevant. We introduce the dimensionless parameter $U$, defined as the ratio of the magic potential depth to the lattice photon recoil energy, $E_r$. Practically speaking, this facilitates replacement of the field amplitudes in Eq.~(\ref{Eq:Vfull}) with trap depth. This reparameterization is convenient, as it is the trap depth that is readily accessible in our experiment (at the fraction of a percent level).

We proceed to regard the magic potential as a zeroth order approximation, subsequently treating $\delta V\left(\rho,z\right)\equiv V\left(\rho,z\right)-V_0\left(\rho,z\right)$ as a perturbation. The zeroth order energies $E^{(0)}_\mathbf{n}$ are independent of the clock state. The first order corrections are given by
\begin{equation*}
E^{(1)}_\mathbf{n}=\left\langle\mathbf{n}\right|\delta V\left(\rho,z\right)\left|\mathbf{n}\right\rangle.
\end{equation*}
In practice, we are interested in $\Delta E^{(1)}_\mathbf{n}$, where $\Delta$ denotes a difference taken between excited state and ground state quantities. As the states $\left|\mathbf{n}\right\rangle$ themselves are state-independent (being defined in terms of the magic potential), we have
\begin{equation*}
\Delta E^{(1)}_\mathbf{n}=\left\langle\mathbf{n}\right|\Delta V\left(\rho,z\right)\left|\mathbf{n}\right\rangle.
\end{equation*}
With Eq.~(\ref{Eq:Vfull}),
\begin{align*}
\Delta E^{(1)}_\mathbf{n}=&{}
-\left(\frac{\mathcal{E}_0}{2}\right)^2\Delta\alpha_{E1}\,X_\mathbf{n}
\\&
-\left(\frac{\mathcal{E}_0}{2}\right)^2\Delta\alpha_{M1E2}\,Y_\mathbf{n}
\\&
-\left(\frac{\mathcal{E}_0}{2}\right)^4\Delta\beta\,Z_\mathbf{n},
\end{align*}
where we've introduced the spatial averages
\begin{gather*}
X_\mathbf{n}\equiv\left\langle\mathbf{n}\left|\exp\left(-\frac{2\rho^2}{w_0^2}\right)\cos^2\left(kz\right)\right|\mathbf{n}\right\rangle,
\\
Y_\mathbf{n}\equiv\left\langle\mathbf{n}\left|\exp\left(-\frac{2\rho^2}{w_0^2}\right)\sin^2\left(kz\right)\right|\mathbf{n}\right\rangle,
\\
Z_\mathbf{n}\equiv\left\langle\mathbf{n}\left|\exp\left(-\frac{4\rho^2}{w_0^2}\right)\cos^4\left(kz\right)\right|\mathbf{n}\right\rangle.
\end{gather*}
Further using the definition of $U$ introduced above and scaling to $h\nu_\mathrm{clock}$, we have
\begin{align*}
\frac{\Delta E^{(1)}_\mathbf{n}}{h\nu_\mathrm{clock}}=&{}
-U\left(\frac{\Delta\alpha_{E1}}{\alpha_{E1}\left(\nu_\mathrm{magic}\right)}\frac{E_r}{h\nu_\mathrm{clock}}\right)X_\mathbf{n}
\\&
-U\left(\frac{\Delta\alpha_{M1E2}}{\alpha_{E1}\left(\nu_\mathrm{magic}\right)}\frac{E_r}{h\nu_\mathrm{clock}}\right)Y_\mathbf{n}
\\&
-U^2\left(\frac{\Delta\beta}{\alpha_{E1}\left(\nu_\mathrm{magic}\right)^2}\frac{E_r^2}{h\nu_\mathrm{clock}}\right)Z_\mathbf{n}.
\end{align*}
The factors appearing in parenthesis here are equivalent to the dimensionless factors $\Delta\alpha_{E1}^\prime$, $\Delta\alpha_{M1E2}^\prime$, and $\Delta\beta^\prime$ of the main text. This is essentially Eq.~(1) of the main text. Contributions higher order in the non-magic effects (e.g., $\Delta E^{(2)}_\mathbf{n}$) are negligible.

One can follow different paths to approximate the factors $X_\mathbf{n}$, $Y_\mathbf{n}$, and $Z_\mathbf{n}$. Briefly, we outline one approach. The full potential, Eq.~(\ref{Eq:Vfull}), is first expanded about the potential minimum at the origin. In the absence of anharmonic effects (terms $\propto\rho^pz^q$ with $p+q\geq2$), the motional states and energies are known exactly. We use these harmonic oscillator states to evaluate the leading anharmonic energy corrections (including terms $\propto\rho^pz^q$ with $p+q=4$) at first order. The resulting expression for the energies is not linear in the state-dependent atomic factors, prompting us to expand to first order in the non-magic parameters $\alpha_{E1}-\alpha_{E1}\left(\nu_\mathrm{magic}\right)$, $\alpha_{M1E2}$, and $\beta$. Finally, taking the difference between clock states yields an expression for $\Delta E_\mathbf{n}$ linear in $\Delta\alpha_{E1}$, $\Delta\alpha_{M1E2}$, and $\Delta\beta$, which is what is desired. The approach outlined here reproduces Eq.~(2) of the main text. Notably, it captures important anharmonic effects of the states $\left|\mathbf{n}\right\rangle$ entering $X_\mathbf{n}$, $Y_\mathbf{n}$, and $Z_\mathbf{n}$.

\subsection{Lattice shift model error}
As described in the main text, we experimentally observe a linear scaling of atomic temperature with trap depth.
Because of this, and for the observed ratio of atomic temperature to trap depth, the thermally-averaged motional quantities in Eq.~(2) exhibit the following approximate relations: $\langle n_{1}\rangle,\langle n_{2}\rangle\!\propto\!\sqrt{U}$ and $\langle n_{3}\rangle,\langle n_{4}\rangle,\langle n_{5}\rangle\!\propto\!U$.
Consequently, Eq.~(2) can be reduced to Eq.~(3), enabling lattice light shifts to be characterized with only linear and quadratic dependencies on trap depth.
These thermally averaged linear~$\alpha^{*}$ and quadratic~$\beta^{*}$ coefficients can be related to the atomic polarizabilities as:
\begin{multline}
\setcounter{equation}{3}
\alpha^{*}=\Delta \alpha_{E1}' \left(\frac{\langle n_{1}\rangle+\langle n_{2}\rangle}{\sqrt{U}}-1\right) \\
+\Delta \alpha_{M1E2}' \left(\frac{\langle n_{5}\rangle}{U}-\frac{\langle n_{1}\rangle}{\sqrt{U}}\right),
\end{multline}
 and
\begin{multline}
\beta^{*}=\Delta\beta' \bigg[1-\frac{2(\langle n_{1}\rangle+\langle n_{2}\rangle)}{\sqrt{U}} \\
+\frac{\langle n_{3}\rangle+\langle n_{4}\rangle+4\langle n_{5}\rangle}{U}\bigg].
\end{multline}
Under these conditions, no additional $U$-dependent fitting terms are required for $10^{-18}$ clock accuracy.

A number of situations may arise where the linear and quadratic terms of Eq.~(3) are not sufficient to accurately model all lattice light shifts.  We consider first the case where atomic temperature does not scale as a simple linear function of trap depth.  Indeed, analysis of motional sideband spectra from our experiment suggests that while the atomic temperature in the transverse dimensions varies predominantly as a linear function of $U$, a small residual quadratic dependence on $U$ is also present.  This higher-order dependence motivates the inclusion of a lattice light shift scaling as $U^3$ in Eq.~(3) of the main text.  Theoretical data produced from a thermal model of Eq.~(2) suggest that fitting with a cubic term is required for correct determination of the light shift at the $1 \times 10^{-18}$ accuracy level.

We note that the importance of accounting for higher-order $U$-dependencies is not primarily motivated by the size of these higher-order light shifts at normal operating conditions, where they are typically small.  Instead, their inclusion is important for correct determination of the lattice laser frequency corresponding to zero linear shift (i.e., $\nu_{\mathrm{zero}} \simeq \nu_{\mathrm{magic}}$), which typically relies on shift measurements made over a range of $U$ including larger values of $U$.  This is important because an error in the determination of $\nu_{\mathrm{zero}}$ can lead to substantial error in the lattice light shift determination for normal operating conditions.  Indeed, it is interesting to note how the determination of $\nu_{\mathrm{zero}}$ changes depending on the number of $U$-dependent terms included in the fit. For example, were the data in Fig.~2 of the main text fit to a purely linear fit (potentially relevant in the absence of hyperpolarizability), then we would deduce $\nu_{\mathrm{zero}} = 394,798,295.6(1.2)$ MHz.  A fit including both linear and quadratic $U$-dependent terms would yield $\nu_{\mathrm{zero}} = 394,798,272.7(1.1)$ MHz.  As indicated in the main text, inclusion of a cubic term yields $\nu_{\mathrm{zero}} = 394,798,267(1)$ MHz.  In this case, we note that the other fitted parameters are: ${\partial\alpha^{*}/\partial\nu_l = 2.46(10)\times 10^{-20} \frac{1}{\mathrm{MHz}}}$, $\beta^{*} = -5.5(2)\times 10^{-22}$, and a cubic term $\gamma^{*} = 9(7) \times 10^{-26}$. Finally, without good physical motivation, we also consider the inclusion of both a cubic and quartic term in the fit of the data, giving $\nu_{\mathrm{zero}} = 394,798,268.8(1.1)$ MHz, essentially consistent with the cubic fit.

Even when temperature remains strictly proportional to  $U$, ultracold temperatures can lead to additional $U$ dependencies of the light shift.
Consider the motional quantity $\langle n_{1}\rangle$.  At the sideband-cooled longitudinal temperatures reported in the main text, the following approximation holds: $\langle n_{1}\rangle \approx \frac{1}{2}\frac{e^{hf_z/k_bT}+1}{e^{hf_z/k_bT}-1}$, where $f_z$, which depends on $U$, is the longitudinal harmonic trap frequency and $k_b$ is the Boltzmann constant.  Taylor expansion of $\langle n_{1}\rangle$ in decreasing orders of $U$ yields several potentially-important terms that give light shifts scaling neither linear nor quadratic in $U$.  At even colder temperatures, $\langle n_{1}\rangle = \langle n_{z}+1/2 \rangle \cong 1/2$, and inspection of Eq. (2) in the main text reveals shifts scaling as $U^{\frac{1}{2}}$ and $U^{\frac{3}{2}}$.  Even at $\nu_{\mathrm{magic}}$, where $\Delta \alpha_{E1}' = 0 $, a shift scaling as $\Delta \alpha_{M1E2}'$ $U^{\frac{1}{2}}$ may challenge the quadratic fit model. For the value of $\Delta\alpha_{M1E2}'$ computed in the main text, the error from a quadratic fit model is $<1 \times 10^{-18}$.  Nevertheless, a fit to theoretical data produced from a thermal model of Eq.~(2) suggests that errors of $>10^{-18}$ can result for cases where $\Delta\alpha_{M1E2}'$ is five times larger.

We note also that, even though experimental details and techniques vary among lattice clock experiments, the ratio of the typical atomic temperature to $U$ often reported in the literature (e.g.~\cite{INRIM2016,Nemitz2016,Ushijima2015cryogenic,bloomNature2014optical,falke2014strontium}) is similar to that measured in the present work.
While the variation of $T$ over a \textit{range} of $U$ has not been reported elsewhere, if $T \propto U$ in other systems, the analysis presented here can be employed with high model accuracy.

\subsection{Lattice spectral purity}
Lattice lasers have a background spectrum of amplified spontaneous emission (ASE)~\cite{siegman86}.
This ASE background can be time-dependent, causing a light shift that introduces a fluctuating error in the determination of $\nu_{\mathrm{magic}}$ and $\nu_{\mathrm{clock}}$.
These effects are pronounced in diode-amplified laser systems~\cite{SyrteTAshifts2012}, whose background spectrum can be just 25~dB below carrier.
Ti:Sapphire lasers exhibit a much lower background spectrum ($>$60~dB below carrier), and are used in the measurements reported here.
The Ti:Sapphire output was spectrally filtered using volume Bragg gratings of various bandwidths (1~nm to 50~pm).
To place experimental bounds on AC Stark shifts due to ASE, we repeat measurements of $\delta\nu_{\mathrm{clock}}/\nu_{\mathrm{clock}}$ using two independent lattice lasers with measurably distinct background spectra.
One system was a traditional Ti:Sapphire laser, while the other was a Ti:Sapphire amplifier seeded by an external cavity diode laser~\cite{Bergeson2002InjTisaph}.  
Using the 50 pm bandpass filter above, we confirmed that both laser systems gave consistent determinations of $\nu_\mathrm{zero}$, with a statistics-limited agreement at $\leq3$ MHz level.

\end{document}